\documentclass[ 
reprint,
amsmath,amssymb,
aps,
]{revtex4-1}

\pdfoutput=1

\usepackage{graphicx}
\usepackage{dcolumn}
\usepackage{bm}

\begin{document}

\preprint{APS/123-QED}

\title{Molecular Dynamics Simulations of Janus Particle Dynamics in Uniform Flow}

\author{Aurelien Y. M. Archereau}
\email{aurelien.archereau@ens.fr}
\affiliation{%
Ecole Normale Superieure and
Pierre and Marie Curie University, France \\
}%
\author{Shaun C. Hendy}
\email{shaun.hendy@auckland.ac.nz}
\affiliation{%
Te P\={u}naha Matatini, Department of Physics, University of Auckland, New Zealand\\
}%
\author{Geoff R. Willmott}
\email{g.willmott@auckland.ac.nz}
\affiliation{%
The MacDiarmid Institute for Advanced Materials and Nanotechnology, Departments of Physics and Chemistry, University of Auckland, New Zealand\\
}%




\date{\today}

\begin{abstract}
We use molecular dynamics simulations to study the dynamics of Janus particles, micro- or nanoparticles which are not spherically symmetric, in the uniform flow of a simple liquid. In particular we consider spheres with an asymmetry in the solid-liquid interaction over their surfaces and calculate the forces and torques experienced by the particles as a function of their orientation with respect to the flow. We also examine particles that are deformed slightly from a spherical shape. We compare the simulation results to the predictions of a previously introduced theoretical approach, which computes the forces and torques on particles with variable slip lengths or aspherical deformations that are much smaller than the particle radius. We find that there is good agreement between the forces and torques computed from our simulations and the theoretical predictions, when the slip condition is applied to the first layer of liquid molecules adjacent to the surface.

\end{abstract}

\maketitle


\section{\label{sec:level1}Introduction}
 
Janus particles \cite{491,719}, named after the Roman god with two faces, are micro- or nanoparticles which are not spherically symmetric. Although many types of asymmetry are possible \cite{722}, the typical example is a spherical particle consisting of two hemispheres with distinct surface properties. Many different types of Janus particles have been reported \cite{491,719}, and possible applications often draw upon the interesting surface wettability, self-assembly, or multifunctional properties of these particles. Multifunctional Janus particles are being used to control affinity for human endothelial cells \cite{application1} or breast cancer cells \cite{application2} in biomedical applications. Examples of the importance of wettability are provided by the use of amphiphilic Janus particles for the stabilization of water-in-oil or oil-in-water emulsions \cite{491}. Much as amphiphilic molecules can form various patterned micelles and vesicles in water \cite{israelachvili1976theory}, collections of Janus particles can also self-assemble into similar phases \cite{1470,1472}. 
 
Links between surface wettability and micro- to nanoscale fluid dynamics have become of significant interest in recent times, particularly in relation to a slip boundary condition \cite{303,335,1057,502,shaun2}. At the solid-liquid interface, the usual hydrodynamic assumption is the Dirichlet (non-slip) boundary condition, in which the fluid adjacent to the surface has zero speed. When this assumption is violated, slip occurs, and non-zero slip has been observed experimentally for atomically smooth, hydrophobic surfaces \cite{303,1057}. The non-slip boundary condition is most commonly described in the rest frame of the wall using the slip length $b$, defined according to Navier's approach \cite{335,455,400,shaun3} in which the velocity ($\mathbf{v}$) of an incompressible fluid has a component tangential to the surface ($v_\parallel$) that is proportional to the shear rate: 
\begin{equation}\label{eq:Slip defn}
v_\parallel=b\mathbf{n}\cdot\left(\nabla\mathbf{v}+\left(\nabla\mathbf{v}\right)^T\right)\cdot\left(\mathbf{I}-\mathbf{nn}\right).
\end{equation}
Here $\mathbf{n}$ is the unit normal to the surface. In a simple shear flow over a planar surface, the slip length represents the depth at which a linear extrapolation of the tangential flow velocity at the surface becomes zero.
 
A Janus particle may have an asymmetric hydrodynamic boundary condition, because different areas on the particle surface can have different slip lengths. Such an asymmetry can generate interesting dynamics, and the force and torque on a slip-asymmetric Janus particle in uniform Newtonian flow have been analytically calculated at the continuum level \cite{geoff1, geoff2}. 
These calculations suggest that a Janus particle consisting of two hemispheres with differing boundary conditions will experience a torque, and have a reduced linear drag coefficient compared to a sphere with a no-slip boundary condition \cite{geoff1}. 
The analytical results apply to particles of radius $R$ in the limit $b \ll  R$, in which case the slip asymmetry is mathematically equivalent to slight deformation of a sphere \cite{463}.
 
Slip-induced dynamics of Janus particles are of practical interest for several reasons. Firstly, experimental evidence for molecular scale slip has been obtained using a limited range of techniques \cite{303,1057,725}. The most successful of these (AFM colloidal probe, and surface force apparatus) measure the force between two surfaces due to drainage, rather than directly measuring the shear force couple between the fluid and the surface. Therefore, studies of slip on micro- and nanospheres could verify and further elucidate the physical nature of molecular scale slip. Secondly, the dynamics provide a mechanism for manipulation and orientation of particles, without the application of external fields, suggesting possible applications as multifunctional particles. Thirdly, the collective dynamics of Janus particles could be affected. Simulations of Janus particle self-assembly \cite{1470,1471,1472} have not so far considered slip.
 
This paper presents molecular dynamics (MD) simulations of Janus particles with asymmetric wettability (and consequently an asymmetry in the slip length) in a simple fluid flow. The values of the force and torque on the particles are calculated as a function of the slip length of the two hemispheres, and the angle between the asymmetry and the flow. Results are compared with the continuum theoretical studies of the slip-asymmetric Janus sphere, and of a slightly deformed sphere with homogeneous wettability. Presently, MD simulations are essential for studying slip-asymmetric Janus particles, because the predicted dynamics have not been experimentally confirmed, mostly due to competition with Brownian motion \cite{geoff1}. Here we make explicit predictions of the forces and torques on particles, whereas previous studies have been largely descriptive \cite{article6,article5}.

\section{\label{sec:level1}Theoretical Approach}

\subsection{\label{sec:level2}The Janus particle}

\begin{figure}[h]
\centering
\includegraphics[width=1.0\linewidth]{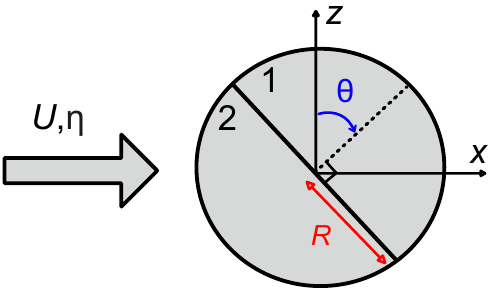}
\caption{Geometry of the system : cross-section through the centre of a Janus sphere in the $x$-$z$ plane. The sphere of radius $R$ is divided into two hemispheres $1$ and $2$. The unbounded uniform flow $U$ is parallel to the $x$ axis, and the sphere is rotated by an angle $\theta$ about the $y$ axis.}
\label{fig:geometry_2}
\end{figure}

The system studied is a Janus particle modelled as a sphere of radius $R$ evenly divided into hemispheres referred to as $1$ and $2$, with the respective slip lengths $b_1$ and $b_2$. The fluid (viscosity $\eta$) moves at a velocity $U$ relative to the Janus particle. The angle $\mathbf{\theta}$ between the sphere and the flow is defined in Fig. \ref{fig:geometry_2}.

In general, the effective slip length measured for flow around a sphere $b^{\parallel}$ will not be equal to the effective slip length measured for flow over a plane $b^{\star}$ for the same liquid-solid interaction strength. As shown by Chen et al. \cite{Chen}, when Eq.~\ref{eq:Slip defn} is used to define slip, for a surface with radius of curvature $R$ we have the relation:
\begin{equation}
\frac{1}{b^{\parallel}}=\frac{1}{b^{\star}}+\frac{1}{R}
\label{Chen}
\end{equation}
where $b^{\parallel} \leq b^{\star}$. Note that the sign of the curvature $R$ differs from some previous accounts \cite{455,geoff1,geoff2} because we define the radius of curvature of the sphere to be positive (and therefore equal to the sphere radius).

We consider the drag force acting parallel to the incident flow, which is in the positive $x$ direction, and the torque about the $y$ axis. If $U$ has a component in the $y$ direction, there are other force and torque components not considered here. It was previously found \cite{geoff1} that the force on a Janus particle with asymmetric slip can be expanded as a first order approximation to the well-known Stokes drag force:
\begin{equation}
F_{Stokes}=6 \pi \eta R U.
\label{Stokes}
\end{equation}
For this Janus particle (Fig.~\ref{fig:geometry_2}), the force and torque are given by

\begin{equation}
\left\{
\begin{aligned}
F_{Janus}=6 \pi \eta R U (1- \frac{b_1^{\parallel}+b_2^{\parallel}}{2R})  \\
T_{Janus}=-9  \pi \eta R^2 U \frac{(b_1^{\parallel}-b_2^{\parallel})}{2R} \cos(\theta) \\
\end{aligned}
\right.
\label{eq_b}
\end{equation}
where $b_1^{\parallel} \ll  R$, $b_2^{\parallel} \ll  R$.

\subsection{\label{sec:level2}The slightly deformed sphere}

We also consider a slightly deformed sphere where the deformation is aligned with the slip asymmetry.  The force and torque on a slightly deformed sphere \cite{463} can be determined using the spherical polar co-ordinate angles $\theta$ (see Fig. \ref{fig:geometry_2}) and $\phi$  to define a point on the surface of the sphere which we associate with the radius $R_d(\theta,\phi)$. The sphere surface takes the locus of points:
\begin{equation}
R_{d}(\theta,\phi)=R (1 + \epsilon  f(\theta,\phi)) 
\end{equation}
where $\epsilon$ is the amplitude of the deformation on one hemisphere ($\epsilon \ll  1$), $\gamma\epsilon$ is the amplitude of the deformation on the other hemisphere ($\epsilon\gamma \ll  1$), and $f(\theta,\phi)=\frac{1}{2}(1+\gamma) + \frac{3}{4}(\gamma-1)cos(\theta)$  is the first order spherical harmonic expansion describing the geometry. 

Calculation of force and torque on the slightly deformed sphere with a non-slip boundary condition leads to the set of equations 
\begin{equation}
\left\{
\begin{aligned}
F_{Janus}=6 \pi \eta R U (1- \epsilon/2(1+\gamma)) \\
T_{Janus}=-\frac{9}{2}  \pi \eta R^2 U \epsilon (1 - \gamma) cos(\theta) \\
\end{aligned}
\label{eq_f}
\right..
\end{equation}
The effects of asymmetric slip (Eq.~\ref{eq_b}) and deformation (Eq.~\ref{eq_f}) are both first order corrections to the Stokes drag (Eq.~\ref{Stokes}), so we can add them to describe a deformed particle with slip asymmetry, leading to the set of equations:
\begin{equation}
\left\{
\begin{aligned}
F_{Janus}=6 \pi \eta R U (1- \epsilon/2(1+\gamma) - \frac{b_1^{\parallel}+b_2^{\parallel}}{2R}) \\
T_{Janus}=-\frac{9}{2}  \pi \eta R^2 U (\epsilon (1 - \gamma)+\frac{b_1^{\parallel}-b_2^{\parallel}}{2R}) cos(\theta) \\
\end{aligned}
\right..
\end{equation}

By choosing a deformed sphere with the same slip condition $b_1^{\parallel}=b_2^{\parallel}=b^{\parallel}$ on both hemispheres, the slip-induced torque vanishes, while the first order correction remains for the force equation:
\begin{equation}
\left\{
\begin{aligned}
F_{Janus}=6 \pi \eta R U (1- \frac{b^{\parallel}}{R} - \epsilon/2(1+\gamma)) \\
T_{Janus}=-\frac{9}{2}  \pi \eta R^2 U \epsilon (1 - \gamma) cos(\theta) \\
\end{aligned}
\right..
\label{eq_def0}
\end{equation}

\section{\label{sec:level1}Methods and Simulations}

\subsection{\label{sec:level2}Inter-molecular interactions}

The molecular dynamics simulations were carried out using an explicit solvent with inter-molecular interactions described by a Lennard-Jones pair potential. For example, the interactions between two fluid monomers (\textit{i},\textit{j}) follow the equation:
\begin{equation}
U_{ij}(r)=4 E_0 [ (\frac{\sigma}{r})^{12} - (\frac{\sigma}{r})^{6}) ] \\
\end{equation}
where $E_0$ represents the depth of the potential well, $r$ is the distance between the centres of the monomers, and $\sigma$ is the distance where $U_{ij}=0$. The potential is truncated at 2.5 $\sigma$ for computational efficiency, at which separation $U_{ij}(2.5\sigma)= 0.016 E_0$.

The surface of the particle is divided evenly in two hemispheres referred to as 1 and 2, each interacting with the fluid monomers by a different Lennard Jones potential, with a different potential well. For example, the interaction between the monomers of hemisphere 1 and the fluid monomers follow the equation:
\begin{equation}
\left\{
\begin{aligned}
U_{ij}(r)=4 E_1 [ (\frac{\sigma}{r})^{12} - (\frac{\sigma}{r})^{6}) ] \\
E_1=A_1E_0
\end{aligned}
\right.
\end{equation}
where $A_1$ is a parameter allowing adjustments of the strength of the pair interaction. As we will see below, the fluid-Janus monomer pair interaction is related to the slip length in that hemisphere, so the slip length $b_1^{\parallel}$ can be controlled by altering $A_1$. The Janus monomers of hemispheres 1 and 2 do not interact with each other.

\subsection{\label{sec:level2}The Janus Particle}
\begin{figure}[ht]
\centering
\includegraphics[width=1.0\linewidth,natwidth=610,natheight=642]{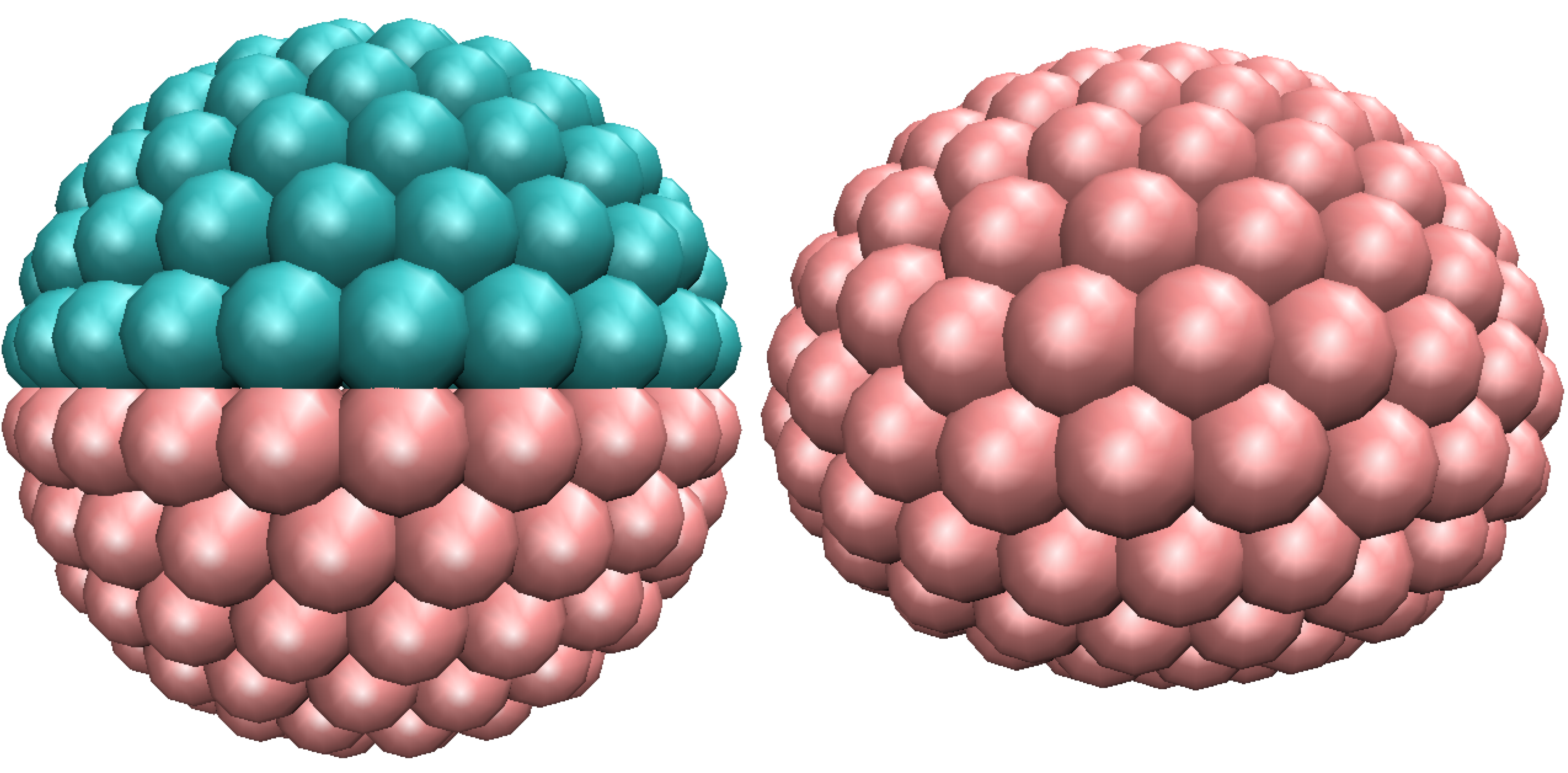}
\caption{Snapshot of the Janus particle (left) and the deformed sphere (right). Both particles are made of 270 Lennard Jones monomers. Note that the deformation of the sphere has been exaggerated here for the sake of visibility.}
\label{janus_pic}
\end{figure}

The Janus particle was constructed from 270 monomers arranged on the surface of a sphere (Fig. \ref{janus_pic}, left). The radius of a sphere containing the monomers is $4.08\sigma$. The theoretical predictions (above) are expected to be valid when $b \ll  R$, and our simulations extend to the upper bound of where we might expect this theory to apply. 

The particle behaves as a rigid body : the relative distance from each monomer on the Janus particle is fixed, and the Janus particle is static in the flow (no translational nor angular motion). The positions of the monomers on the Janus particle were calculated using a Monte Carlo Scheme so that the monomers were evenly distributed on the surface. Each monomer (\textit{i},\textit{j}) on the surface is defined by the two angles ($\phi_1^i$,$\phi_2^i$). We designed an arbitrary score function which strongly disadvantages configurations in which two monomers are close, defined as
\begin{equation}
S=\sum_{i\neq j} ( \frac{1}{d_{ij}} )^4,
\end{equation}
where (\textit{i},\textit{j}) refer to the monomers of the Janus particle. The score function can be seen as a repulsive force between close monomers. Convergence was assessed by studying the first peak of the pair distribution function. A new Monte Carlo design of the Janus particle was done before each simulation in order to account for the irregularities of the particle.

The deformed homegeneous sphere  (Fig.~\ref{janus_pic}, right) was designed using the same Monte Carlo scheme, with a uniform pair interaction potential ($A_1=A_2$).

\subsection{\label{sec:level2}Molecular dynamics simulations}

The equations of motion are integrated using a Verlet method with the timestep set at $\Delta t=0.005 \tau_{LJ}$, where $\tau_{LJ}$ is the Lennard-Jones timescale $\left(m_0 \sigma/E_0\right)^{1/2}$ and $m_0$ is the monomer mass. All simulations are run in Lennard-Jones units, where $\sigma$ is the unit of distance, $E_0$ is the unit of energy, $m_0$ the unit of mass and $\tau_{LJ}$ the unit of time, without loss of generality. The simulations were carried out with LAMMPS \cite{lammps1} in the canonical ensemble (constant NVT). The temperature was set at $1.1 \epsilon / k_B$ (where $k_B$ is the Boltzmann constant) using a Langevin thermostat with a damping factor (coupling parameter) set at 20 $\tau_{LJ}$. Initial velocities for the monomers were drawn from a Maxwell-Boltzmann distribution. 

For the study of the Janus particle, the system consisted of $46 288$ fluid monomers of mass $m_0$ in a cubic box with sides of length 39.62$\sigma$ ($\rho_{fluid}=0.75\sigma^{-3}$). Periodic boundary conditions are used in the three dimensions. The flow is created by moving the Janus particle at a constant speed ($U=0.3\sigma/\tau$) in a static fluid. The effect of the finite box size was assessed by monitoring the average fluid speed at the boundary of the box, which was found to be less than 3\% of the speed of the Janus particle in all simulations reported here.

We computed the fluid's shear viscosity $\eta$ using the Green-Kubo relation, which relates the viscosity to the autocorrelation function of a diagonal component of the pressure tensor. This was calculated using a system with only fluid monomers in a box with periodic boundary conditions and with the fluid at the same fluid density as the fluid in the Janus particle simulations. The calculation converged in $2\times10^7$ MD steps to a value of $\eta=3.40 \pm 0.01$ $m_0/\sigma \tau_{LJ}$.

Slip lengths $b_P^{\star}$ for flow over a planar surface were computed as a function of the fluid-Janus particle interaction energy $E=A E_0$ by simulating a Poiseuille flow between two solid walls in the $yz$ plane. The walls were composed of (111) planes of face-centered cubic lattice (i.e. a close-packed surface) with a surface density corresponding to that of the Janus particle. The Poiseuille flow was created by adding a constant force $F_x=0.005 E_0 / \sigma$ to all the fluid monomers. The velocity profile across the channel was then fitted with a parabolic curve to determine the slip length at the walls for each value of the fluid-wall monomer interaction energy $E=A E_0$. The exact position of the walls in this calculation is discussed further below.

The number of monomers in our structures varied between around 10 000 and 60 000. Simulations were allowed $2\times 10^5$ time steps for equilibration, followed by $8\times 10^5$ time steps for production runs. For each type of simulation 10 simulations were run to estimate the mean and standard deviation of computed quantities.

\section{\label{sec:level1}Results}

\subsection{\label{sec:level2}Effect of thermostat on the slip length}
\begin{figure}[h]
\centering
\includegraphics[width=1\linewidth]{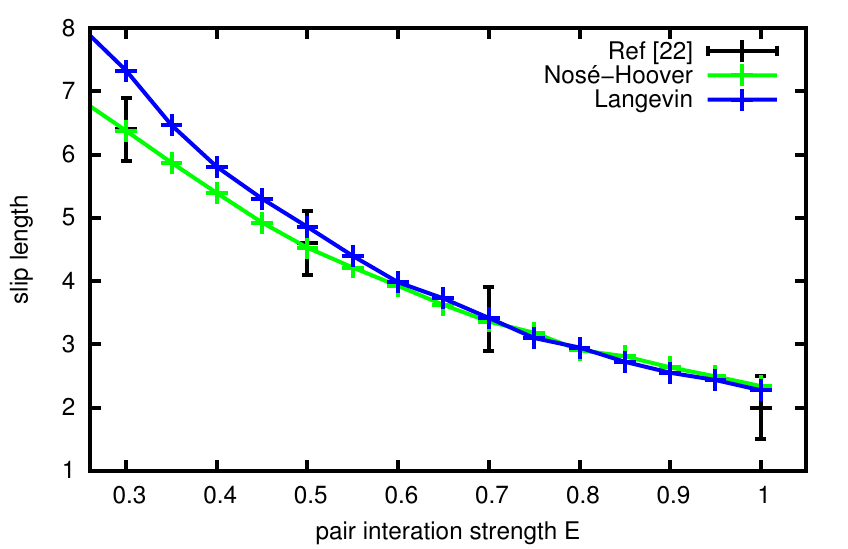}
\caption{Slip length in $\sigma$ units, as a function of $E=A E_0$ on the surface. Literature values of $b_P^{\star}$ from Ref \cite{article6} are shown in black, we were able to replicate these results (green). The result with the Langevin thermostat is shown in blue.}
\label{slip1}
\end{figure}

Although, as discussed above, we use a Langevin thermostat for most of the work reported here, it is important to check whether this has an impact on computed slip lengths. Figure~\ref{slip1} shows the slip length for a planar surface $b_P^{\star}$ as a function of $E=A E_0$ for both the Langevin thermostat (Fig.~\ref{slip1}, blue line), a Nos\'{e}-Hoover thermostat (Fig.~\ref{slip1}, green line) and several values from Ref~\cite{article6}. The Nos\'{e}-Hoover thermostat used a coupling parameter of 1.0 $\tau_{LJ}$, which is the same value used in Ref~\cite{article6}. The differences are relatively small, but our choice of thermostat enabled us to avoid the `flying ice cube' effect \cite{flying} over a wider range of simulation parameters. This choice of thermostat leads to slightly larger values of slip length for small $E$ than the simulations in Ref~\cite{article6}.

\subsection{\label{sec:level2}Slip for a homogeneous sphere}

\begin{figure}[h]
\centering
\includegraphics[width=1\linewidth]{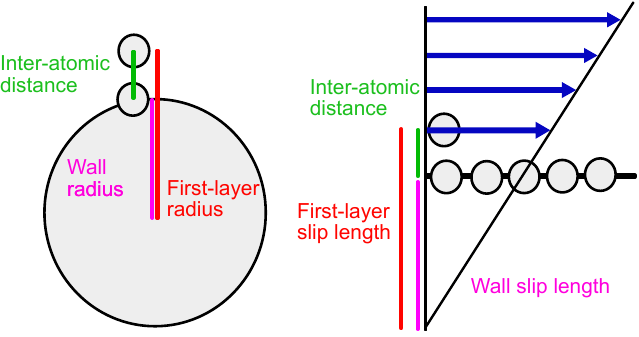}
\caption{Both the effective radius $R$ of the simulated sphere (left) and the slip length (right) can be defined using either the position of the solid wall (surface), or the first layer of fluid monomers. The latter definition was consistently used here.}
\label{slip3}
\end{figure}

\begin{figure}[h]
\centering
\includegraphics[width=1\linewidth]{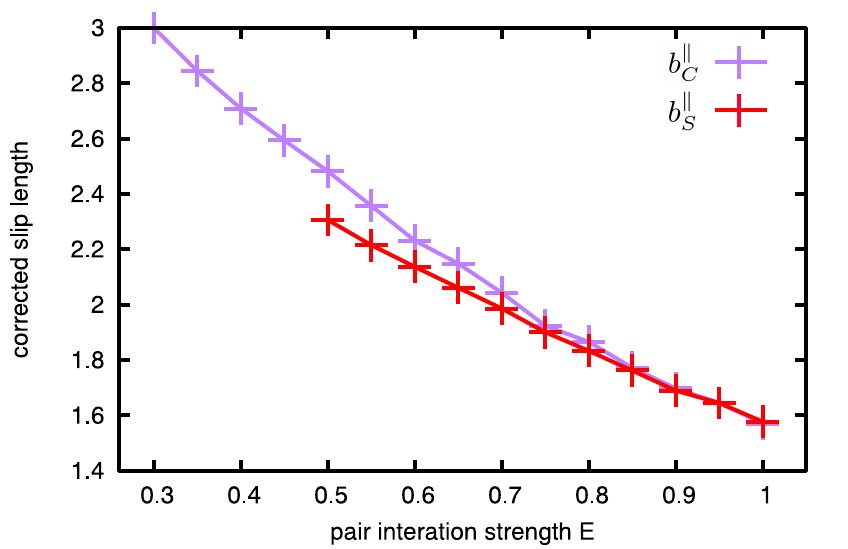}
\caption{The corrected values of the slip lengths $b_C^{\parallel}$ (purple), along with the slip length values $b_S^{\parallel}$ computed from the simulation of the homogeneous sphere (red).}
\label{slip2}
\end{figure}

We can compute the slip length for a homogeneous sphere in order to check the applicability of Eq.~\ref{Chen}. Using Eq. \ref{eq_b} with $b_1^{\parallel}=b_2^{\parallel}$, an effective slip length $b_S^{\parallel}$ can be extracted from the measured force on a simulated homogeneous sphere for various values of the interaction strength $E$: 
\begin{equation}
b_S^{\parallel}=R(1-\frac{F}{6 \pi \eta R U})
\end{equation}
The calculations to find the Stokes-extracted slip length $b_S^{\parallel}$ and the corrected slip length $b_C^{\parallel}$ both use the radius of the sphere $R$. There are two different, but reasonable definitions for the radius of the sphere (see Fig. \ref{slip3}) : the distance from the center of the sphere to the surface monomers ($R=4.08$ $\sigma$), or the distance from the center of the sphere to the first layer of fluid ($R=5.08$ $\sigma$). The difference is 20\%, and therefore not negligible. 

Analysis of the simulations shows that the latter definition has the better agreement, and is used in what follows throughout. In particular, we find that the slip length calculated from the homogeneous particle $b^{\parallel}_S$ is in good agreement with the values calculated from the Poiseuille flow if the effective radius is chosen to be $R=5.08$ $\sigma$ (see Fig. \ref{slip2}). This makes physical sense, as it is in the first layer of fluid where the boundary condition (Eq.~\ref{eq:Slip defn}) is defined. At low slip length the agreement is especially good whereas at high slip length a difference is visible, which is consistent with the fact that the theory used the theoretical assumption $b \ll R$. This is not the case here, as the maximum slip length is nearly half the radius. In the following, the slip length used to calculate the expected force and torque is the slip length $b_S^{\parallel}$ extracted from the homogeneous sphere simulations.

\subsection{\label{sec:level2}Force and torque for a Janus particle}

\begin{figure*}[ht]\centering
\includegraphics[width=1\linewidth,natwidth=610,natheight=642]{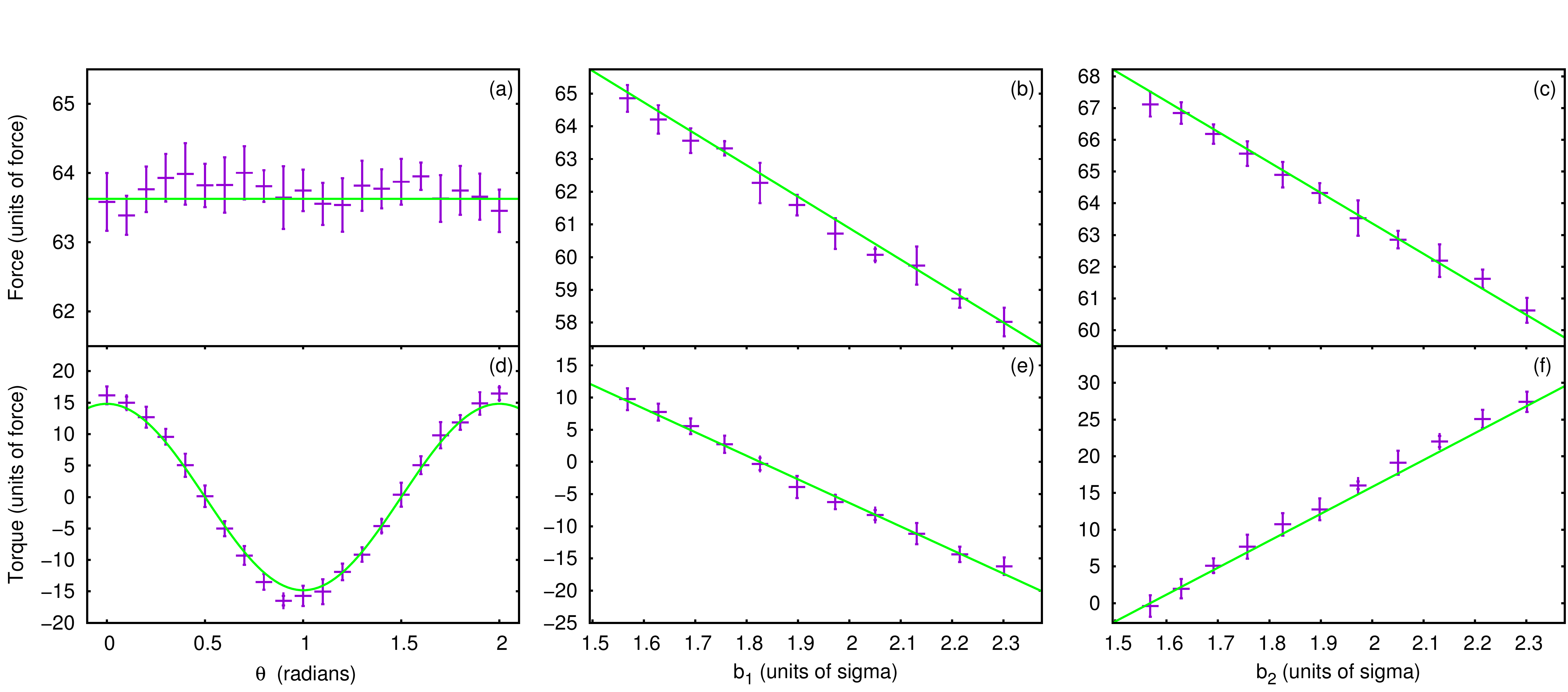}
\caption{Force and torque values as a function of $\theta$, $b_1^{\parallel}$ and $b_2^{\parallel}$ for the Janus particle simulation. The values calculated from simulations are plotted in purple, and the expected values (Eq.~\ref{eq_b}) are shown in green. Fixed parameters : for a,d : $A_1=1.0$, $A_2=0.7$. For b,e : $A_2=0.8$, $\theta=0$. For c,f : $A_1=1.0$, $\theta=0$.}
\label{S_1}
\end{figure*}

To assess the agreement of the simulations with Eqs.~\ref{eq_b}, we calculated the force and torque on simulated Janus particles with two different values of the interaction strength $R$ as a function of the angle $\theta$, and the two slip lengths $b_1^{\parallel}$ and $b_2^{\parallel}$ imputed from the calculations in the previous section. The results, shown in Fig.~\ref{S_1}, are in good agreement with the the expected values from the theory. The functional agreement is high (the fits have Pearson coefficients larger than 0.994) and the magnitude of the simulated values are always within 4\% of the expected ones.

One can detect a small variation of frequency $\pi$ in the subplot \textbf{(a)}, representing a variation in the force as a function of the angle. This variation is likely to be the consequence of a second order term, as the values of slip length are of the same order of magnitude as the radius of the sphere. One other possibility is the design of the sphere, which is not regular along the equator between the two poles (see Fig. \ref{janus_pic}). The amplitude of this effect is, however, small : 0.27\% of the force value. We conclude that the theory does an excellent job of describing the forces and torques on our simulated Janus particle.

\subsection{\label{sec:level2}Force and torque for a deformed sphere}

\begin{figure*}[ht]\centering
\includegraphics[width=1.0\linewidth,natwidth=610,natheight=642]{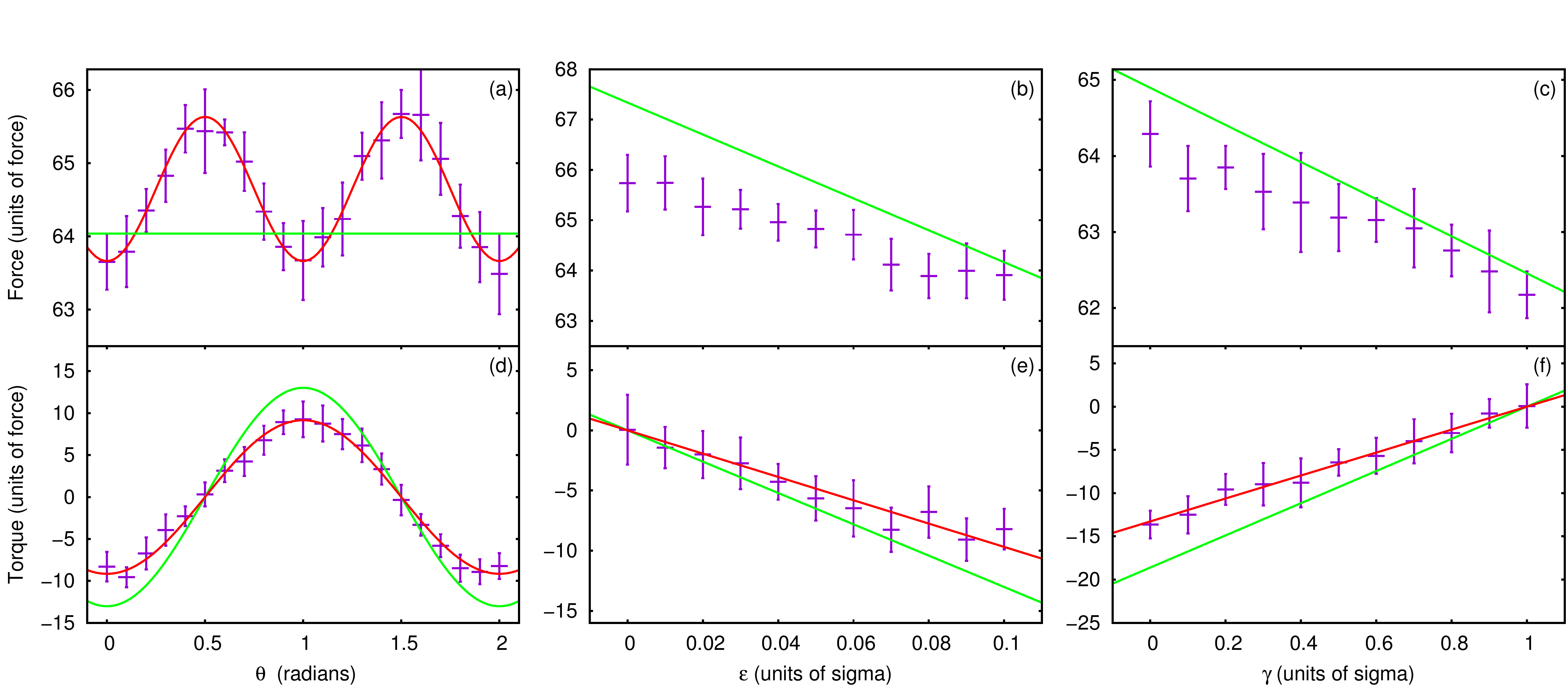}
\caption{Force and torque values as a function of $\theta$, $\epsilon$ and $\gamma$ for the deformed sphere simulation. The values are plotted in purple, and the expected values (Eqs.~\ref{eq_def0}) are shown in green. On subplot \textbf{(a)} a fit with a cosine is presented and shows a second-order effect (red). On subplots \textbf{(d,e,f)} a fit is shown in red. Fixed parameters : for a,d : $\epsilon=0.1$, $\gamma=0.3$. For b,e : $\theta=0$, $\gamma=0.3$. For c,f : $\theta=0$, $\epsilon=0.1$.}
\label{D_1}
\end{figure*}

Finally we calculated the force and torque using simulations for a deformed sphere, which should be described by Eqs.~\ref{eq_def0}, as a function of the angle $\theta$, and the deformation parameters $\epsilon$ and $\gamma$ (see Fig. \ref{D_1}). We designed a deformed sphere with homogeneous interaction strength $E$ where the slip length $b^{\parallel}$ is as small as possible ($A=1.0$ everywhere, thus we expect that $b^{\parallel}=1.58 \sigma$). Again, we find that the simulations are in good agreement with the expected values. The differences are larger than in the case of the Janus particle simulations, quite possibly due to the combination of the slip and the deformation.

In particular, the simulation of the force as a function of the angle (subplot \textbf{(a)}) exhibits a sinusoidal deviation of frequency $\pi$ from the expected behavior. Again, this might be a second order effect as in the Janus particle set of simulations, or a combined effect of the slip and the deformation. The amplitude of this effect is however small : less than 2\% of the value of the force. A theoretical analysis to include second order terms is not trivial. 

However, when fitting the data with a free multiplicative parameter $B$ as follows 
\begin{equation}
T=-B*\frac{9}{2}  \pi \eta R^2 U \epsilon (1 - \gamma) cos(\theta) \\
\label{eq_def0}
\end{equation}
we find that $B=0.72$, which is of the order of $(1-\frac{b^{\parallel}}{R})=0.69$ in these simulations. This correction can be physically interpreted as the first order torque acting from an adjusted distance $R_{adjusted}=R(1-\frac{b^{\parallel}}{R})$.

\section{\label{sec:level1}Conclusion}
For small Janus particles in a simple liquid flow, we find good agreement between the forces and torques computed from molecular dynamics simulations and a continuum theoretical description based on inhomogeneous perturbations to a Stokes flow about a homogenous sphere. We have considered particles with an asymmetry in the solid-liquid interaction strength over their surfaces and particles that are deformed slightly from a spherical shape but have a uniform interaction strength. The theory gives excellent quantitative agreement with the simulations, particularly in the limit of small perturbations or slip lengths as used in the theory, and describes the functional dependence of forces and torques on angle well. The results demonstrate the importance of applying the slip condition at the first layer of fluid molecules, and of adjusting the slip length to account for surface curvature. This work suggests that molecular dynamics simulations could be used to study related problems with more complicated geometries and flows.

\section*{Acknowledgments}
The molecular dynamics simulations were run using the LAMMPS software (Large-Scale Molecular Massively Parallel Simulator \cite{lammps1}). The calculations were run on NeSI Pan Cluster, part of the Centre for eResearch hosted by the University of Auckland. 

\bibliography{Janus}

\begin{thebibliography}{26}%
\makeatletter
\providecommand \@ifxundefined [1]{%
 \@ifx{#1\undefined}
}%
\providecommand \@ifnum [1]{%
 \ifnum #1\expandafter \@firstoftwo
 \else \expandafter \@secondoftwo
 \fi
}%
\providecommand \@ifx [1]{%
 \ifx #1\expandafter \@firstoftwo
 \else \expandafter \@secondoftwo
 \fi
}%
\providecommand \natexlab [1]{#1}%
\providecommand \enquote  [1]{``#1''}%
\providecommand \bibnamefont  [1]{#1}%
\providecommand \bibfnamefont [1]{#1}%
\providecommand \citenamefont [1]{#1}%
\providecommand \href@noop [0]{\@secondoftwo}%
\providecommand \href [0]{\begingroup \@sanitize@url \@href}%
\providecommand \@href[1]{\@@startlink{#1}\@@href}%
\providecommand \@@href[1]{\endgroup#1\@@endlink}%
\providecommand \@sanitize@url [0]{\catcode `\\12\catcode `\$12\catcode
  `\&12\catcode `\#12\catcode `\^12\catcode `\_12\catcode `\%12\relax}%
\providecommand \@@startlink[1]{}%
\providecommand \@@endlink[0]{}%
\providecommand \url  [0]{\begingroup\@sanitize@url \@url }%
\providecommand \@url [1]{\endgroup\@href {#1}{\urlprefix }}%
\providecommand \urlprefix  [0]{URL }%
\providecommand \Eprint [0]{\href }%
\providecommand \doibase [0]{http://dx.doi.org/}%
\providecommand \selectlanguage [0]{\@gobble}%
\providecommand \bibinfo  [0]{\@secondoftwo}%
\providecommand \bibfield  [0]{\@secondoftwo}%
\providecommand \translation [1]{[#1]}%
\providecommand \BibitemOpen [0]{}%
\providecommand \bibitemStop [0]{}%
\providecommand \bibitemNoStop [0]{.\EOS\space}%
\providecommand \EOS [0]{\spacefactor3000\relax}%
\providecommand \BibitemShut  [1]{\csname bibitem#1\endcsname}%
\let\auto@bib@innerbib\@empty
\bibitem [{\citenamefont {Perro}\ \emph {et~al.}(2005)\citenamefont {Perro},
  \citenamefont {Reculusa}, \citenamefont {Ravaine}, \citenamefont
  {Bourgeat-Lami},\ and\ \citenamefont {Duguet}}]{491}%
  \BibitemOpen
  \bibfield  {author} {\bibinfo {author} {\bibfnamefont {A.}~\bibnamefont
  {Perro}}, \bibinfo {author} {\bibfnamefont {S.}~\bibnamefont {Reculusa}},
  \bibinfo {author} {\bibfnamefont {S.}~\bibnamefont {Ravaine}}, \bibinfo
  {author} {\bibfnamefont {E.}~\bibnamefont {Bourgeat-Lami}}, \ and\ \bibinfo
  {author} {\bibfnamefont {E.}~\bibnamefont {Duguet}},\ }\href@noop {}
  {\bibfield  {journal} {\bibinfo  {journal} {J. Mater. Chem.}\ }\textbf
  {\bibinfo {volume} {15}},\ \bibinfo {pages} {3745} (\bibinfo {year}
  {2005})}\BibitemShut {NoStop}%
\bibitem [{\citenamefont {Walther}\ and\ \citenamefont {Muller}(2008)}]{719}%
  \BibitemOpen
  \bibfield  {author} {\bibinfo {author} {\bibfnamefont {A.}~\bibnamefont
  {Walther}}\ and\ \bibinfo {author} {\bibfnamefont {A.~H.~E.}\ \bibnamefont
  {Muller}},\ }\href@noop {} {\bibfield  {journal} {\bibinfo  {journal} {Soft
  Matter}\ }\textbf {\bibinfo {volume} {4}},\ \bibinfo {pages} {663} (\bibinfo
  {year} {2008})}\BibitemShut {NoStop}%
\bibitem [{\citenamefont {Glotzer}\ and\ \citenamefont {Solomon}(2007)}]{722}%
  \BibitemOpen
  \bibfield  {author} {\bibinfo {author} {\bibfnamefont {S.~C.}\ \bibnamefont
  {Glotzer}}\ and\ \bibinfo {author} {\bibfnamefont {M.~J.}\ \bibnamefont
  {Solomon}},\ }\href@noop {} {\bibfield  {journal} {\bibinfo  {journal}
  {Nature Materials}\ }\textbf {\bibinfo {volume} {6}},\ \bibinfo {pages} {557}
  (\bibinfo {year} {2007})}\BibitemShut {NoStop}%
\bibitem [{\citenamefont {Yoshida}\ \emph {et~al.}(2009)\citenamefont
  {Yoshida}, \citenamefont {Roh}, \citenamefont {Mandal}, \citenamefont
  {Bhaskar}, \citenamefont {Lim}, \citenamefont {Nandivada}, \citenamefont
  {Deng},\ and\ \citenamefont {Lahann}}]{application1}%
  \BibitemOpen
  \bibfield  {author} {\bibinfo {author} {\bibfnamefont {M.}~\bibnamefont
  {Yoshida}}, \bibinfo {author} {\bibfnamefont {K.-H.}\ \bibnamefont {Roh}},
  \bibinfo {author} {\bibfnamefont {S.}~\bibnamefont {Mandal}}, \bibinfo
  {author} {\bibfnamefont {S.}~\bibnamefont {Bhaskar}}, \bibinfo {author}
  {\bibfnamefont {D.}~\bibnamefont {Lim}}, \bibinfo {author} {\bibfnamefont
  {H.}~\bibnamefont {Nandivada}}, \bibinfo {author} {\bibfnamefont
  {X.}~\bibnamefont {Deng}}, \ and\ \bibinfo {author} {\bibfnamefont
  {J.}~\bibnamefont {Lahann}},\ }\href@noop {} {\bibfield  {journal} {\bibinfo
  {journal} {Advanced materials}\ }\textbf {\bibinfo {volume} {21}},\ \bibinfo
  {pages} {4920} (\bibinfo {year} {2009})}\BibitemShut {NoStop}%
\bibitem [{\citenamefont {Wu}\ \emph {et~al.}(2010)\citenamefont {Wu},
  \citenamefont {Ross}, \citenamefont {Hong},\ and\ \citenamefont
  {Lee}}]{application2}%
  \BibitemOpen
  \bibfield  {author} {\bibinfo {author} {\bibfnamefont {L.~Y.}\ \bibnamefont
  {Wu}}, \bibinfo {author} {\bibfnamefont {B.~M.}\ \bibnamefont {Ross}},
  \bibinfo {author} {\bibfnamefont {S.}~\bibnamefont {Hong}}, \ and\ \bibinfo
  {author} {\bibfnamefont {L.~P.}\ \bibnamefont {Lee}},\ }\href@noop {}
  {\bibfield  {journal} {\bibinfo  {journal} {Small}\ }\textbf {\bibinfo
  {volume} {6}},\ \bibinfo {pages} {503} (\bibinfo {year} {2010})}\BibitemShut
  {NoStop}%
\bibitem [{\citenamefont {Israelachvili}\ \emph {et~al.}(1976)\citenamefont
  {Israelachvili}, \citenamefont {Mitchell},\ and\ \citenamefont
  {Ninham}}]{israelachvili1976theory}%
  \BibitemOpen
  \bibfield  {author} {\bibinfo {author} {\bibfnamefont {J.~N.}\ \bibnamefont
  {Israelachvili}}, \bibinfo {author} {\bibfnamefont {D.~J.}\ \bibnamefont
  {Mitchell}}, \ and\ \bibinfo {author} {\bibfnamefont {B.~W.}\ \bibnamefont
  {Ninham}},\ }\href@noop {} {\bibfield  {journal} {\bibinfo  {journal}
  {Journal of the Chemical Society, Faraday Transactions 2: Molecular and
  Chemical Physics}\ }\textbf {\bibinfo {volume} {72}},\ \bibinfo {pages}
  {1525} (\bibinfo {year} {1976})}\BibitemShut {NoStop}%
\bibitem [{\citenamefont {Rosenthal}\ \emph {et~al.}(2012)\citenamefont
  {Rosenthal}, \citenamefont {Gubbins},\ and\ \citenamefont {Klapp}}]{1470}%
  \BibitemOpen
  \bibfield  {author} {\bibinfo {author} {\bibfnamefont {G.}~\bibnamefont
  {Rosenthal}}, \bibinfo {author} {\bibfnamefont {K.~E.}\ \bibnamefont
  {Gubbins}}, \ and\ \bibinfo {author} {\bibfnamefont {S.~H.~L.}\ \bibnamefont
  {Klapp}},\ }\href {\doibase 10.1063/1.4707954} {\bibfield  {journal}
  {\bibinfo  {journal} {The Journal of Chemical Physics}\ }\textbf {\bibinfo
  {volume} {136}},\ \bibinfo {pages} {174901} (\bibinfo {year}
  {2012})}\BibitemShut {NoStop}%
\bibitem [{\citenamefont {Li}\ \emph {et~al.}(2012)\citenamefont {Li},
  \citenamefont {Lu}, \citenamefont {Sun},\ and\ \citenamefont {An}}]{1472}%
  \BibitemOpen
  \bibfield  {author} {\bibinfo {author} {\bibfnamefont {Z.~W.}\ \bibnamefont
  {Li}}, \bibinfo {author} {\bibfnamefont {Z.~Y.}\ \bibnamefont {Lu}}, \bibinfo
  {author} {\bibfnamefont {Z.-Y.}\ \bibnamefont {Sun}}, \ and\ \bibinfo
  {author} {\bibfnamefont {L.~J.}\ \bibnamefont {An}},\ }\href {\doibase
  10.1039/c2sm25397f} {\bibfield  {journal} {\bibinfo  {journal} {Soft Matter}\
  }\textbf {\bibinfo {volume} {8}},\ \bibinfo {pages} {6693} (\bibinfo {year}
  {2012})}\BibitemShut {NoStop}%
\bibitem [{\citenamefont {Neto}\ \emph {et~al.}(2005)\citenamefont {Neto},
  \citenamefont {Evans}, \citenamefont {Bonaccurso}, \citenamefont {Butt},\
  and\ \citenamefont {Craig}}]{303}%
  \BibitemOpen
  \bibfield  {author} {\bibinfo {author} {\bibfnamefont {C.}~\bibnamefont
  {Neto}}, \bibinfo {author} {\bibfnamefont {D.~R.}\ \bibnamefont {Evans}},
  \bibinfo {author} {\bibfnamefont {E.}~\bibnamefont {Bonaccurso}}, \bibinfo
  {author} {\bibfnamefont {H.-J.}\ \bibnamefont {Butt}}, \ and\ \bibinfo
  {author} {\bibfnamefont {V.~S.~J.}\ \bibnamefont {Craig}},\ }\href@noop {}
  {\bibfield  {journal} {\bibinfo  {journal} {Rep. Prog. Phys.}\ }\textbf
  {\bibinfo {volume} {68}},\ \bibinfo {pages} {2859} (\bibinfo {year}
  {2005})}\BibitemShut {NoStop}%
\bibitem [{\citenamefont {Lauga}\ \emph {et~al.}(2005)\citenamefont {Lauga},
  \citenamefont {Brenner},\ and\ \citenamefont {Stone}}]{335}%
  \BibitemOpen
  \bibfield  {author} {\bibinfo {author} {\bibfnamefont {E.}~\bibnamefont
  {Lauga}}, \bibinfo {author} {\bibfnamefont {M.~P.}\ \bibnamefont {Brenner}},
  \ and\ \bibinfo {author} {\bibfnamefont {H.~A.}\ \bibnamefont {Stone}},\
  }\enquote {\bibinfo {title} {Handbook of experimental fluid dynamics},}\ \
  (\bibinfo  {publisher} {Springer},\ \bibinfo {address} {New York},\ \bibinfo
  {year} {2005})\ Chap.\ \bibinfo {chapter} {15: Microfluidics: The No-Slip
  Boundary Condition}\BibitemShut {NoStop}%
\bibitem [{\citenamefont {Bocquet}\ and\ \citenamefont
  {Charlaix}(2010)}]{1057}%
  \BibitemOpen
  \bibfield  {author} {\bibinfo {author} {\bibfnamefont {L.}~\bibnamefont
  {Bocquet}}\ and\ \bibinfo {author} {\bibfnamefont {E.}~\bibnamefont
  {Charlaix}},\ }\href {\doibase 10.1039/b909366b} {\bibfield  {journal}
  {\bibinfo  {journal} {Chem. Soc. Rev.}\ }\textbf {\bibinfo {volume} {39}},\
  \bibinfo {pages} {1073} (\bibinfo {year} {2010})}\BibitemShut {NoStop}%
\bibitem [{\citenamefont {Bocquet}\ and\ \citenamefont {Barrat}(2007)}]{502}%
  \BibitemOpen
  \bibfield  {author} {\bibinfo {author} {\bibfnamefont {L.}~\bibnamefont
  {Bocquet}}\ and\ \bibinfo {author} {\bibfnamefont {J.-L.}\ \bibnamefont
  {Barrat}},\ }\href@noop {} {\bibfield  {journal} {\bibinfo  {journal} {Soft
  Matter}\ }\textbf {\bibinfo {volume} {3}},\ \bibinfo {pages} {685} (\bibinfo
  {year} {2007})}\BibitemShut {NoStop}%
\bibitem [{\citenamefont {Hendy}\ and\ \citenamefont {Lund}(2007)}]{shaun2}%
  \BibitemOpen
  \bibfield  {author} {\bibinfo {author} {\bibfnamefont {S.~C.}\ \bibnamefont
  {Hendy}}\ and\ \bibinfo {author} {\bibfnamefont {N.~J.}\ \bibnamefont
  {Lund}},\ }\href {\doibase 10.1103/PhysRevE.76.066313} {\bibfield  {journal}
  {\bibinfo  {journal} {Phys. Rev. E}\ }\textbf {\bibinfo {volume} {76}},\
  \bibinfo {pages} {066313} (\bibinfo {year} {2007})}\BibitemShut {NoStop}%
\bibitem [{\citenamefont {Einzel}\ \emph {et~al.}(1990)\citenamefont {Einzel},
  \citenamefont {Panzer},\ and\ \citenamefont {Liu}}]{455}%
  \BibitemOpen
  \bibfield  {author} {\bibinfo {author} {\bibfnamefont {D.}~\bibnamefont
  {Einzel}}, \bibinfo {author} {\bibfnamefont {P.}~\bibnamefont {Panzer}}, \
  and\ \bibinfo {author} {\bibfnamefont {M.}~\bibnamefont {Liu}},\ }\href@noop
  {} {\bibfield  {journal} {\bibinfo  {journal} {Phys. Rev. Lett.}\ }\textbf
  {\bibinfo {volume} {64}},\ \bibinfo {pages} {2269} (\bibinfo {year}
  {1990})}\BibitemShut {NoStop}%
\bibitem [{\citenamefont {Navier}(1827)}]{400}%
  \BibitemOpen
  \bibfield  {author} {\bibinfo {author} {\bibfnamefont {C.~L. M.~H.}\
  \bibnamefont {Navier}},\ }\href@noop {} {\bibfield  {journal} {\bibinfo
  {journal} {Mem. Acad. Sci. Inst. Fr.}\ }\textbf {\bibinfo {volume} {6}},\
  \bibinfo {pages} {839} (\bibinfo {year} {1827})}\BibitemShut {NoStop}%
\bibitem [{\citenamefont {Lund}\ \emph {et~al.}(2012)\citenamefont {Lund},
  \citenamefont {Zhang}, \citenamefont {Mahelona},\ and\ \citenamefont
  {Hendy}}]{shaun3}%
  \BibitemOpen
  \bibfield  {author} {\bibinfo {author} {\bibfnamefont {N.~J.}\ \bibnamefont
  {Lund}}, \bibinfo {author} {\bibfnamefont {X.~P.}\ \bibnamefont {Zhang}},
  \bibinfo {author} {\bibfnamefont {K.}~\bibnamefont {Mahelona}}, \ and\
  \bibinfo {author} {\bibfnamefont {S.~C.}\ \bibnamefont {Hendy}},\ }\href
  {\doibase 10.1103/PhysRevE.86.046303} {\bibfield  {journal} {\bibinfo
  {journal} {Phys. Rev. E}\ }\textbf {\bibinfo {volume} {86}},\ \bibinfo
  {pages} {046303} (\bibinfo {year} {2012})}\BibitemShut {NoStop}%
\bibitem [{\citenamefont {Willmott}(2008)}]{geoff1}%
  \BibitemOpen
  \bibfield  {author} {\bibinfo {author} {\bibfnamefont {G.~R.}\ \bibnamefont
  {Willmott}},\ }\href@noop {} {\bibfield  {journal} {\bibinfo  {journal}
  {Physical Review E}\ }\textbf {\bibinfo {volume} {77}},\ \bibinfo {pages}
  {055302} (\bibinfo {year} {2008})}\BibitemShut {NoStop}%
\bibitem [{\citenamefont {Willmott}(2009)}]{geoff2}%
  \BibitemOpen
  \bibfield  {author} {\bibinfo {author} {\bibfnamefont {G.~R.}\ \bibnamefont
  {Willmott}},\ }\href@noop {} {\bibfield  {journal} {\bibinfo  {journal}
  {Physical Review E}\ }\textbf {\bibinfo {volume} {79}},\ \bibinfo {pages}
  {066309} (\bibinfo {year} {2009})}\BibitemShut {NoStop}%
\bibitem [{\citenamefont {Happel}\ and\ \citenamefont {Brenner}(1973)}]{463}%
  \BibitemOpen
  \bibfield  {author} {\bibinfo {author} {\bibfnamefont {J.}~\bibnamefont
  {Happel}}\ and\ \bibinfo {author} {\bibfnamefont {H.}~\bibnamefont
  {Brenner}},\ }\href@noop {} {\emph {\bibinfo {title} {Low Reynolds Number
  Hydrodynamics}}}\ (\bibinfo  {publisher} {Noordhoff International
  Publishing},\ \bibinfo {address} {The Netherlands},\ \bibinfo {year}
  {1973})\BibitemShut {NoStop}%
\bibitem [{\citenamefont {Bouzigues}\ \emph {et~al.}(2008)\citenamefont
  {Bouzigues}, \citenamefont {Bocquet}, \citenamefont {Charlaix}, \citenamefont
  {Cottin-Bizonne}, \citenamefont {Cross}, \citenamefont {Joly}, \citenamefont
  {Steinberger}, \citenamefont {Ybert},\ and\ \citenamefont {Tabeling}}]{725}%
  \BibitemOpen
  \bibfield  {author} {\bibinfo {author} {\bibfnamefont {C.~I.}\ \bibnamefont
  {Bouzigues}}, \bibinfo {author} {\bibfnamefont {L.}~\bibnamefont {Bocquet}},
  \bibinfo {author} {\bibfnamefont {E.}~\bibnamefont {Charlaix}}, \bibinfo
  {author} {\bibfnamefont {C.}~\bibnamefont {Cottin-Bizonne}}, \bibinfo
  {author} {\bibfnamefont {B.}~\bibnamefont {Cross}}, \bibinfo {author}
  {\bibfnamefont {L.}~\bibnamefont {Joly}}, \bibinfo {author} {\bibfnamefont
  {A.}~\bibnamefont {Steinberger}}, \bibinfo {author} {\bibfnamefont
  {C.}~\bibnamefont {Ybert}}, \ and\ \bibinfo {author} {\bibfnamefont
  {P.}~\bibnamefont {Tabeling}},\ }\href@noop {} {\bibfield  {journal}
  {\bibinfo  {journal} {Phil. Trans. R. Soc. A}\ }\textbf {\bibinfo {volume}
  {366}},\ \bibinfo {pages} {1455} (\bibinfo {year} {2008})}\BibitemShut
  {NoStop}%
\bibitem [{\citenamefont {Moghani}\ and\ \citenamefont {Khomami}(2013)}]{1471}%
  \BibitemOpen
  \bibfield  {author} {\bibinfo {author} {\bibfnamefont {M.~M.}\ \bibnamefont
  {Moghani}}\ and\ \bibinfo {author} {\bibfnamefont {B.}~\bibnamefont
  {Khomami}},\ }\href {\doibase 10.1039/c3sm27345h} {\bibfield  {journal}
  {\bibinfo  {journal} {Soft Matter, 2013, 9,}\ }\textbf {\bibinfo {volume}
  {9}},\ \bibinfo {pages} {4815} (\bibinfo {year} {2013})}\BibitemShut
  {NoStop}%
\bibitem [{\citenamefont {Kharazmi}\ and\ \citenamefont
  {Priezjev}(2015)}]{article6}%
  \BibitemOpen
  \bibfield  {author} {\bibinfo {author} {\bibfnamefont {A.}~\bibnamefont
  {Kharazmi}}\ and\ \bibinfo {author} {\bibfnamefont {N.~V.}\ \bibnamefont
  {Priezjev}},\ }\href@noop {} {\bibfield  {journal} {\bibinfo  {journal}
  {Journal Of Chemical Physics}\ }\textbf {\bibinfo {volume} {142}},\ \bibinfo
  {pages} {234503} (\bibinfo {year} {2015})}\BibitemShut {NoStop}%
\bibitem [{\citenamefont {Bianchi}\ \emph {et~al.}(2015)\citenamefont
  {Bianchi}, \citenamefont {Panagiotopoulos},\ and\ \citenamefont
  {Nikoubashman}}]{article5}%
  \BibitemOpen
  \bibfield  {author} {\bibinfo {author} {\bibfnamefont {E.}~\bibnamefont
  {Bianchi}}, \bibinfo {author} {\bibfnamefont {A.~Z.}\ \bibnamefont
  {Panagiotopoulos}}, \ and\ \bibinfo {author} {\bibfnamefont {A.}~\bibnamefont
  {Nikoubashman}},\ }\href@noop {} {\bibfield  {journal} {\bibinfo  {journal}
  {Soft matter}\ }\textbf {\bibinfo {volume} {11}},\ \bibinfo {pages} {3767}
  (\bibinfo {year} {2015})}\BibitemShut {NoStop}%
\bibitem [{\citenamefont {Chen}\ \emph {et~al.}(2014)\citenamefont {Chen},
  \citenamefont {Zhang},\ and\ \citenamefont {Koplik}}]{Chen}%
  \BibitemOpen
  \bibfield  {author} {\bibinfo {author} {\bibfnamefont {W.}~\bibnamefont
  {Chen}}, \bibinfo {author} {\bibfnamefont {R.}~\bibnamefont {Zhang}}, \ and\
  \bibinfo {author} {\bibfnamefont {J.}~\bibnamefont {Koplik}},\ }\href
  {\doibase 10.1103/PhysRevE.89.023005} {\bibfield  {journal} {\bibinfo
  {journal} {Phys. Rev. E}\ }\textbf {\bibinfo {volume} {89}},\ \bibinfo
  {pages} {023005} (\bibinfo {year} {2014})}\BibitemShut {NoStop}%
\bibitem [{\citenamefont {Plimpton}(1995)}]{lammps1}%
  \BibitemOpen
  \bibfield  {author} {\bibinfo {author} {\bibfnamefont {S.}~\bibnamefont
  {Plimpton}},\ }\href@noop {} {\bibfield  {journal} {\bibinfo  {journal}
  {Journal of computational physics}\ }\textbf {\bibinfo {volume} {117}},\
  \bibinfo {pages} {1} (\bibinfo {year} {1995})}\BibitemShut {NoStop}%
\bibitem [{\citenamefont {Harvey}\ \emph {et~al.}(1998)\citenamefont {Harvey},
  \citenamefont {Tan},\ and\ \citenamefont {Cheatham}}]{flying}%
  \BibitemOpen
  \bibfield  {author} {\bibinfo {author} {\bibfnamefont {S.~C.}\ \bibnamefont
  {Harvey}}, \bibinfo {author} {\bibfnamefont {R.~K.-Z.}\ \bibnamefont {Tan}},
  \ and\ \bibinfo {author} {\bibfnamefont {T.~E.}\ \bibnamefont {Cheatham}},\
  }\href {\doibase
  10.1002/(SICI)1096-987X(199805)19:7<726::AID-JCC4>3.0.CO;2-S} {\bibfield
  {journal} {\bibinfo  {journal} {Journal of Computational Chemistry}\ }\textbf
  {\bibinfo {volume} {19}},\ \bibinfo {pages} {726} (\bibinfo {year}
  {1998})}\BibitemShut {NoStop}%
\end{thebibliography}%

\end{document}